\documentclass[runningheads]{llncs}
\usepackage[T1]{fontenc}
\usepackage{graphicx}
\usepackage{amsmath,amssymb,amsthm}
\usepackage{upgreek}
\usepackage{thmtools}
\usepackage{nameref}
\usepackage[pdfborder={0 0 0}]{hyperref}
\usepackage[table]{xcolor}
\usepackage{caption}
\usepackage{ifthen}
\usepackage{multirow}
\usepackage{hhline}
\usepackage{siunitx}
\usepackage{booktabs}
\usepackage{longtable}
\usepackage{pdflscape}
\usepackage{rotating}
\usepackage{tablefootnote}
\usepackage{tabularray}\UseTblrLibrary{booktabs}
\usepackage{threeparttable}
\usepackage{makecell}
\usepackage{subcaption}
\usepackage{placeins}
\usepackage{listings}
\usepackage{tikz}
\usepackage{soul}
\usepackage{inline-images}
\usepackage{enumitem}
\usepackage{fancyhdr}

\usepackage{mymacros}

\usetikzlibrary{arrows.meta}

\begin{document}
\title{A Scalability Analysis of Quantitative Confidence Assessment Methods for Assurance Cases}
%
%
\author{Simon Diemert\inst{1}\orcidID{0000-0001-9493-7969} \and
Jens H. Weber\inst{1}\orcidID{0000-0003-4591-6728}}
\authorrunning{S. Diemert and J. H. Weber}
%
\institute{University of Victoria, Victoria, British Columbia, Canada\\
\email{\{sdiemert,jens\}@uvic.ca}
}
\maketitle              
\begin{tikzpicture}[remember picture, overlay]
  \node[anchor=south, yshift=2cm] at (current page.south) {%
    \parbox{0.9\textwidth}{\centering\color{red}\small
      Preprint accepted to the SASSUR'26 workshop. This preprint has not undergone peer review or any
      post-submission improvements or corrections. The Version of Record of this contribution
      is published in the SafeCOMP workshop proceedings when they become available via Springer.
    }
  };
\end{tikzpicture}
\begin{abstract}
This paper proposes a model to estimate the decision complexity and effort required to apply quantitative confidence assessment methods to assurance cases. The model considers both the worst and average case for these measures and characterizes how these quantities scale with argument size. Prior work has indicated that the additional effort required to apply these methods is a barrier to their adoption by assurance case practitioners. Researchers developing new methods, or improving existing methods, can use this model to estimate the effort required to apply their method. The proposed model is parameterized using data from published case studies and is applied to three existing quantitative confidence assessment methods: the Bayesian Belief Network method, the Dempster-Shafer Theory method, and the Certus method. The results show that, while Certus has the highest worst-case decision complexity, its average-case effort is lower than the BBN and DST methods.

\keywords{Assurance cases \and Confidence assessment \and Scalability}
\end{abstract}

\newcommand{\tightSubsectionSpacing}{\vspace{-0.3cm}}
\newcommand{\tightParagraphSpacing}{\vspace{-0.3cm}}

\section{Introduction}\label{sec:intro}

Preparing an assurance case (\AC{}) is a necessary activity for assuring safety- and security-critical systems in numerous domains, such as the automotive industry \cite{iso26262,iso8800}. \AC{}s are structured arguments, supported by evidence, that a system or organisation will satisfy desired quality attributes within a defined operating environment \cite{gsn}. Numerous notations have been developed to express, organize, and manage \AC{}s, with the most widely adopted being Goal Structuring Notation (GSN) \cite{gsn,kelly}. However, the use of a structured notation alone does not ensure that an \AC{} has adequately argued that a system fulfills its quality attribute(s). In particular, the authors or reviewers of an \AC{} may be left wondering: \textit{do we believe the top-level claim is true?}.

Confidence assessment methods (\CAM{}s) offer systematic means to evaluate an \AC{} \cite{diemert2025cams}. Numerous \CAM{}s exist, and may be categorized as: qualitative (e.g., Assurance Claim Points \cite{hawkins2011}, Eliminative Argumentation \cite{goodenough2015}, and iTest \cite{holloway2021}), quantitative (e.g., methods using Bayesian Networks \cite{hobbs2012,nesic2021} or Dempster-Shafer Theory \cite{idmessaoud2024}), or mixed (e.g., Assurance 2.0 \cite{bloomfield2023} or \Certus{} \cite{diemert2025certus}).

A recent survey of practitioners revealed that the additional effort required to apply a \CAM{}, beyond what is normally required when preparing an \AC{}, is a key barrier to adoption \cite{diemert2025cams}. In particular, for quantitative or mixed methods, significant effort might be required to both assign the computation's parameters (e.g., weights on the argument graph's edges) and to provide assessments for each piece of evidence supporting the argument (i.e., leaves in the argument graph). Graydon and Holloway observed a lack of scalability data or analyses for quantitative \CAM{}s, remarking that: ``it is not clear that [the surveyed] techniques requiring a substantial effort for each [argument step] will be feasible in practice ... the selected papers present no empirical evidence of scalability'' \cite{graydon2017}. It follows that researchers developing, or improving, quantitative \CAM{}s, should evaluate the level of effort required to apply the method. 

This paper develops a numerical model for estimating the ``decision complexity'' of a quantitative \CAM{}, i.e., the number of decisions a user of the \CAM{} must make when applying it to an argument of a given size in both the worst case and average use cases. Using this model, the scalability of three quantitative \CAM{}s is evaluated as the argument size changes. In this evaluation, parameters related to argument size are estimated based on previously published \AC{}s. Then, the model is further parameterized to estimate the effort (i.e., time spent by a user) required to apply the method. The methods and results in this workshop paper are based on an analysis that was recently published in the first author's doctoral thesis \cite{diemert2026thesis}. To our knowledge, this paper is the first to address the scalability of quantitative \CAM{}s.

The remainder of this paper is organized as follows. To begin, each of the three \CAM{}s under analysis are introduced in enough detail to justify our modelling decisions. Next, the model for decision complexity and effort is introduced and then applied to the \CAM{}s. The paper closes with a discussion of results, limitations, and next steps.
\section{Methods for Quantitative Confidence Assessment}\label{sec:bg}

The scalability model developed in this paper is applicable to \CAM{}s where: 1) the argument is expressed using a structured notation that represents the argument as a tree-like directed acyclic graph; 2) the user supplies valuations (belief, confidence, etc.) at the leaves of the argument; 3) the user annotates the internal structure of the argument with parameters or logic used to propagate quantities through the argument; and 4) a calculation or algorithm is applied to produce a valuation of argument's top-level claim.

Three \CAM{}s satisfying these criteria are introduced below: the Bayesian Belief Network (BBN) method due to Hobbs and Lloyd \cite{hobbs2012}, the Dempster-Shafer Theory (DST) method developed by Idmessaoud et al. \cite{idmessaoud2024}, and the \Certus{} method developed by Diemert and Weber \cite{diemert2025certus,diemert2026thesis}. While variations on both the BBN and DST methods exist (e.g., \cite{nesic2021}), these methods were chosen as representative methods for their respective underlying theory of uncertainty. For instance, Hobbs and Lloyd's BBN method is founded on probability theory.

\subsection{Overview of the BBN Method}\label{sec:bg-bbn}

Hobbs and Lloyd model an \AC{}'s argument structure using a BBN \cite{hobbs2012}. Each claim in the argument is assigned a subjective degree of belief $\Pr(c_i) \in [0,1]$, where 1 means the claim is believed true and 0 means it is believed false. Each leaf node in the argument, usually evidence, is annotated with a probability indicating the user's belief. The user also provides three parameters for each step in the argument to configure the propagation of belief from child nodes to the parent node:

\begin{itemize}[leftmargin=*,itemsep=4pt]

\item \bb{Combinator}: Either \noisyAND{} or \noisyOR{}, describing whether the parent is supported conjunctively or disjunctively by its children.

\item \bb{Leakage}: Captures the residual uncertainty not explained by the children. For \noisyAND{}, the leakage defines the belief the parent is false even if all children are true, and for \noisyOR{}, it defines the belief the parent is true even if all children are false.

\item \bb{Link Weight}: Defines the influence of a child on the parent's belief. For \noisyAND{}, this is the reduction in belief in the parent if the child alone is false, and for \noisyOR{}, it is the increase in belief if the child alone is true.

\end{itemize}

These parameters are used to calculate the values in a conditional probability table (CPT) for each argument step. Then, following the standard procedure for Bayesian Networks, a conditional probability formula is used to compute belief in the parent from the children, thus propagating belief through the argument.

\subsection{Overview of the DST Method}\label{sec:bg-dst}

Idmessaoud et al. applied Dempster-Shafer Theory (DST), which is a generalization of probability theory \cite{idmessaoud2024}. Their method uses two measures to quantify the belief in the truth of a claim: \term{decision} $d \in \{d_1, \ldots, d_5\}$, which captures the degree of acceptability from fully rejectable ($d_1$) to fully acceptable ($d_5$), and \term{confidence} $c \in \{c_1, \ldots, c_5\}$, which captures the conviction behind the decision.

For each leaf node in the argument, the user supplies both a decision and confidence assessment. Additionally, the user must provide several parameters for each argument step: 

\begin{itemize}[itemsep=4pt]

    \item \bb{Combinator}: either Simple, Conjunction, Disjunction, or Hybrid, indicating how to combine belief in a step's children.

    \item \bb{Direct Parameter}: for each child node, describing the impact of the child on the parent if the child claim is believed to be true.

    \item \bb{Reverse Parameter}: for each child node, describing the impact of the child on the parent if the child claim is believed to be false.

    \item \bb{Direct Rule}: for the parent node, describing the collective impact the children have on the parent, if the children are all true.

    \item \bb{Reverse Rule}: for the parent node, describing the collective impact the children have on the parent, if the children are all false.

\end{itemize}

As with the BBN method above, these parameters are used to configure a series of computations that propagate the user's assessments from the leaves of the argument to the root to produce an overall \textit{decision} and \textit{confidence} measure.

\subsection{Overview of the \Certus{} Method}\label{sec:bg-certus}

\Certus{} is a mixed (quantitative and qualitative) method for confidence assessment that uses a domain-specific language to describe belief in an \AC{}'s argument \cite{diemert2025certus,diemert2026thesis}. Rather than assigning numerical probabilities, \Certus{} uses linguistic belief levels to describe belief in the argument. Nine ``canonical'' belief levels are suggested, ranging from \textit{reject} (the claim is surely false) to \textit{certain} (the claim is surely true). The belief levels represent fuzzy sets over a belief scale defined using possibility theory \cite{diemert2026thesis}. The user may choose to interact with \Certus{} linguistically, using the named belief levels, or numerically, using degrees of membership in fuzzy sets. The levels range from \il{reject} (the claim is surely false) through \il{uncertain} (maximum uncertainty) to \il{certain} (the claim is surely true), providing an expressive yet interpretable vocabulary for belief.

\begin{figure}[t]
    \centering
    \includegraphics[width=0.9\columnwidth]{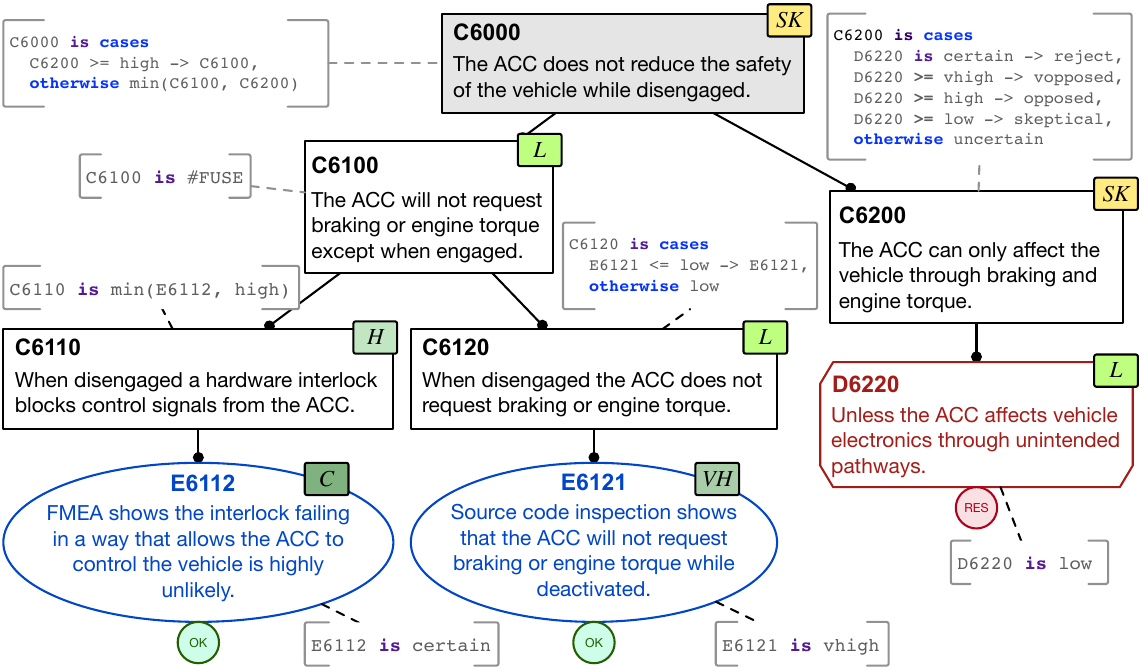}
    \caption{Applying \Certus{} to an argument fragment; from \cite{diemert2026thesis}.}
    \label{fig:certus-intro}
\end{figure}

To use \Certus{}, the user annotates each node in the argument with a belief assignment expression. Leaf nodes are usually assigned belief levels directly, e.g., \il{E1 is certain}. Internal nodes in the argument are annotated with expressions that describe how belief in the children combine to produce a belief in the parent. The \Certus{} language provides several mechanisms for expressing belief propagation: comparison and logical operators (e.g., \il{>=}, \il{and}), built-in functions (e.g., \il{min}, \il{max}), \il{cases} conditional expressions, and higher-level macros such as \il{#MIN} (the minimum among the child beliefs) and \il{#FUSE} (approximately the average belief among child nodes). Figure~\ref{fig:certus-intro} illustrates these ideas on a small adaptive cruise control (ACC) argument\footnotemark from \cite{diemert2026thesis}. Belief annotations appear in the top-right corner of each node (e.g., \texttt{L} for \il{low}, \texttt{SK} for \il{skeptical}), and \Certus{} expressions are shown in partial rectangles connected by dashed lines.

\footnotetext{This argument fragment is intended for illustrative purposes, and might be incompleted. Notably, it does not address the possibility of driver ``mode confusion''.}


\section{Model for Estimating Decision Complexity and Effort}\label{sec:method}

This section develops models for estimating the decision complexity and level of effort required to apply a quantitative \CAM{} for worst case and average case of using the \CAM{}. For the purpose of this analysis, ``decision complexity'' is defined as the number of decisions a user must make while applying a \CAM{}, including decisions related to selecting leaf valuations or selecting propagation parameters. The level of effort model aims to translate the decision complexity into a measure that represents the duration of time a user might require to complete the analysis. The ``worst case'' measure imagines that user must make every possible decision entailed by the method whereas the ``average case'' assumes that sensible default values and tool support can be used to reduce the number of decisions.

\subsection{Modelling an Argument Structure}

\begin{figure}[h]
    \centering
    \includegraphics[width=0.7\textwidth]{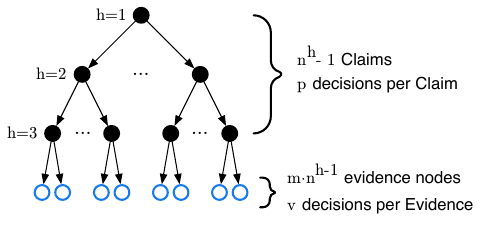}
    \caption{Model of \AC{} argument structure for analysis.}
    \label{fig:scalability-tree}
\end{figure}

For this analysis, an \AC{}'s argument is modelled as an $n$-ary tree of claims with height $h > 1$, where each leaf claim has $m$ child evidence nodes. While real \AC{}s have much more diverse structures, this admittedly artificial model of argument simplifies the analysis while also permitting meaningful comparison among quantitative \CAM{}s. This model is visualized in Figure~\ref{fig:scalability-tree} a tree with $h=3$.

The $n$-ary tree of claims has a total of $n^h - 1$ nodes, each being a parent whose belief is computed based on $n \in \N$ children. Let $p \in \N$ be the number of decisions that a user of a \CAM{} must make for each parent node. The $h^{\text{th}}$ layer of the claim tree has $n^{h-1}$ parent nodes that are each supported by $m \in \N$ evidence nodes. In total there are $m \cdot n^{h-1}$ evidence nodes. Let $v \in \N$ denote the number of decisions that a user must make about each evidence node.

\subsection{Worst Case Decision Complexity}

Using the above construction of an argument, the number of decisions that a user of a \CAM{} must make to apply the method in the \it{worst} case is:
\begin{equation}\label{eq:scalability-worst-case}
O_d\Big(\big[p \cdot (n^h - 1)\big] + \big[v \cdot m \cdot n^{h-1} \big]\Big)
\end{equation}
Where $O(\ldots)$ is Big O notation, the first term is the number of decisions to configure belief propagation among claims, and the second term gives the number of leaf valuations.

\subsection{Average Case Decision Complexity}

In practice, the user does not need to make a decision for every analysis input. Method implementations or tools can use sensible defaults that represent the most common configurations. Let $\alpha_n$ and $\alpha_m$ denote the proportion of cases where a user must fully specify an argument step's configuration ($p$ decisions) or leaf valuation ($v$ decisions). In the other $1 - \alpha$ cases, a smaller number of decisions are required, $p'$ and $v'$ respectively:
\begin{equation}\label{eq:scalability-avg-case}
\Omega_d\Big(
    (n^h - 1) \cdot \big[\alpha_n \cdot p + (1 - \alpha_n) \cdot p' \big]
    +
    m \cdot n^{h-1}\big[\alpha_m \cdot v + (1 - \alpha_m) \cdot v' \big]
    \Big)
\end{equation}
When $\alpha_n = \alpha_m = 1$, the average case reduces to the worst case above.

\subsection{Estimating Effort}

Let $t_p$ denote the time in minutes to make a single decision about belief propagation in an argument step, and $t_v$ the time to make a single leaf valuation decision. The total effort in minutes is:
\begin{equation}\label{eq:scalability-time}
T =
    t_p \cdot (n^h - 1) \cdot \big[\alpha_n \cdot p + (1 - \alpha_n) \cdot p' \big]
    +
    t_v \cdot m \cdot n^{h-1}\big[\alpha_m \cdot v + (1 - \alpha_m) \cdot v' \big]
\end{equation}

\section{Selecting Model Parameters}\label{sec:param}

The model for decision complexity and effort derived in Section \ref{sec:method} has a number of parameters that must be selected before it can be used for a scalability analysis.

\subsection{Selecting Argument Size Parameters}

The two parameters, $n$ and $m$, govern the size of arguments generated by this model. These parameters are estimated using data from a selection of published \AC{} case studies as shown in Table~\ref{tab:argument-sizes}. Using the averages values, summing evidence and residual defeater nodes for leaf count and claims plus defeaters for internal nodes gives:
\begin{align*}
222 = n^h - 1 \quad&\text{and}\quad 131 = m \cdot n^{h-1}
\end{align*}
Taking $h = 10$ yields $n \approx 1.72$ and $m \approx 1$, which is consistent with \AC{}s having 1--2 child claims per argument step and one evidence node per line of reasoning.

\begin{table}[h]
    \centering
    \footnotesize
    \caption{Reported node counts for a selection of \AC{}s in the literature.}
    \label{tab:argument-sizes}
    \renewcommand{\arraystretch}{1.4}
    \begin{tabular}{p{4cm}C{1.1cm}C{1.5cm}C{1.5cm}C{1.3cm}C{0.8cm}}
        \toprule
        \bb{Argument}
            & \bb{Claims}
            & \bb{Evidence}
            & \bb{Defeaters}
            & \bb{Res.}
            & \bb{Sum}
        \\\midrule
        CERN LHC MPS \cite{viger2025cern}
            & 146 & 70  & 105 & 9  & 330 \\\lightrule
        Argument \#1 from \cite{diemert2020}
            & 184 & 147 & 153 & 19 & 503 \\\lightrule
        Argument \#2 from \cite{diemert2020}
            & 188 & 144 & 130 & 28 & 490 \\\lightrule
        Argument \#3 from \cite{diemert2020}
            & 94  & 81  & 79  & 7  & 261 \\\lightrule
        ILI Trust Argument \cite{casey2024ipc}
            & 142 & 165 & 45  & -  & 352 \\\lightrule
        Ventilator Argument \cite{gvent}
            & 113 & 54  & 81  & 37 & 285 \\\midrule
        \bb{Average}
            & 144 & 110 & 99  & 21 & 370 \\\bottomrule
    \end{tabular}
    
\end{table}

\subsection{Selecting $\alpha$ for the Average Case Analysis}
For the average case, we select $\alpha_n = \alpha_m = 0.20$. This corresponds to a scenario where 1 in 5 argument steps requires non-trivial decision-making. This value was informed by observations from other analyses, including a case study and expressivity analysis, in which 21\% of reasoning steps required custom belief propagation expressions \cite{diemert2026thesis}.

\subsection{Selecting \CAM{}-Specific Parameters}

The decision complexity and effort models require that the $p$, $p'$, $v$, and $v'$ parameters be selected for each \CAM{} being analyzed.

\tightParagraphSpacing{}
\paragraph{BBN Method.} In the worst case decision complexity, a user of the BBN method must make $p = n + 2$ decisions per parent claim in the argument. Each child has a link weight parameter and then the combinator and leakage parameter must be selected. For evidence a single decision is made per node, so $v = 1$. For the average decision complexity, let the $\alpha$ case be the same as the worst case above ($p = n + 2$, $v = 1$). However, suppose that for the $1 - \alpha$ case, tooling assists the user so that they must only select the combinator (\noisyAND{} or \noisyOR{}) for each parent claim. Further, suppose the tool also assigns link weights for each child to evenly distribute their contribution to the parent (e.g., $w_i = 1 / n$ for \noisyAND{}). The leakage parameter defaults to $k = 1$, giving $p' = 1$.  For evidence, the user is still required to assign a belief for each leaf node, so $v' = 1$.

\tightParagraphSpacing{}
\paragraph{DST Method.} For the worst case decision complexity, a user of the DST method must make $p = 2n + 3$ decisions per parent claim in the argument. Each child has two parameters (the forward and reverse parameter) and each parent has a combinator, a combining forward, and combining reverse parameter. For evidence, two decisions must be provided per node, one for the ``decision'' and ``confidence'', so $v = 2$. For the average decision complexity, let the $\alpha$ case be the same as the worst case above ($p = 2n + 3$, $v = 2$). However, suppose that for the $1 - \alpha$ case tooling assists the user so that they only need to select a combinator, so $p' = 1$. For the evidence, two decisions must still be made, so $v' = 2$.

\tightParagraphSpacing{}
\paragraph{\Certus{}.} For the worst case decision complexity, suppose that a user of \Certus{} must write a custom belief propagation expression over all child nodes in an argument step and belief levels. For an argument step \step{$c_0$}{$c_1, c_2$}, this is: 

\begin{lstlisting}[language=certus]
    c0 is cases
        c1 is reject and c2 is reject -> reject,
        ...
        c1 is certain and c2 is vhigh -> vhigh,
        c1 is certain and c2 is certain -> certain,
        otherwise uncertain
\end{lstlisting}

There are nine belief levels defined by \Certus{}, so this gives $p = 9^n$ decisions per argument step for the worst case. The number of leaf valuations for evidence is $v = 1$, since a user can make an assignment with a single decision. This worst case for \Certus{} is quite extreme; several features of the \Certus{} language exist to avoid such scenarios, including macros and user-defined operators. For the average case, suppose a user must manually craft a belief propagation expression with $p = n$ decisions $\alpha$ percent of the time, and in the other $1 - \alpha$ cases they can use \Certus{}' macros or a user-defined operator that require only a single decision, so $p' = 1$. Valuation of leaf nodes remains as $v' = 1$ since every leaf must be assigned a belief level as input.

\subsection{Selecting Time and Effort Parameters}

The parameters $t_p$ and $t_v$ must be chosen to model effort/time estimates. For simplicity, the same parameters are used in both the worst and average case analyses. Suppose that each evidence valuation takes (on average) $t_v = 5$ minutes of effort to complete. In practice, some might be much faster and others might take significantly longer. For example, many evidence checks amount to simple binary decisions (e.g., ``was this document approved?''), which could be completed within a few seconds, but others might require lengthy reviews, consultation with interest holders, and so on. For the argument steps, suppose that each decision takes $t_p = 0.5$ minutes (30 seconds) to complete. As with evidence valuation, this is intended as an average, with many decisions taking just a few seconds and others requiring more time.

\section{Scalability Analysis}\label{sec:results}

\begin{figure}[h]
    \centering
    \includegraphics[width=0.9\textwidth]{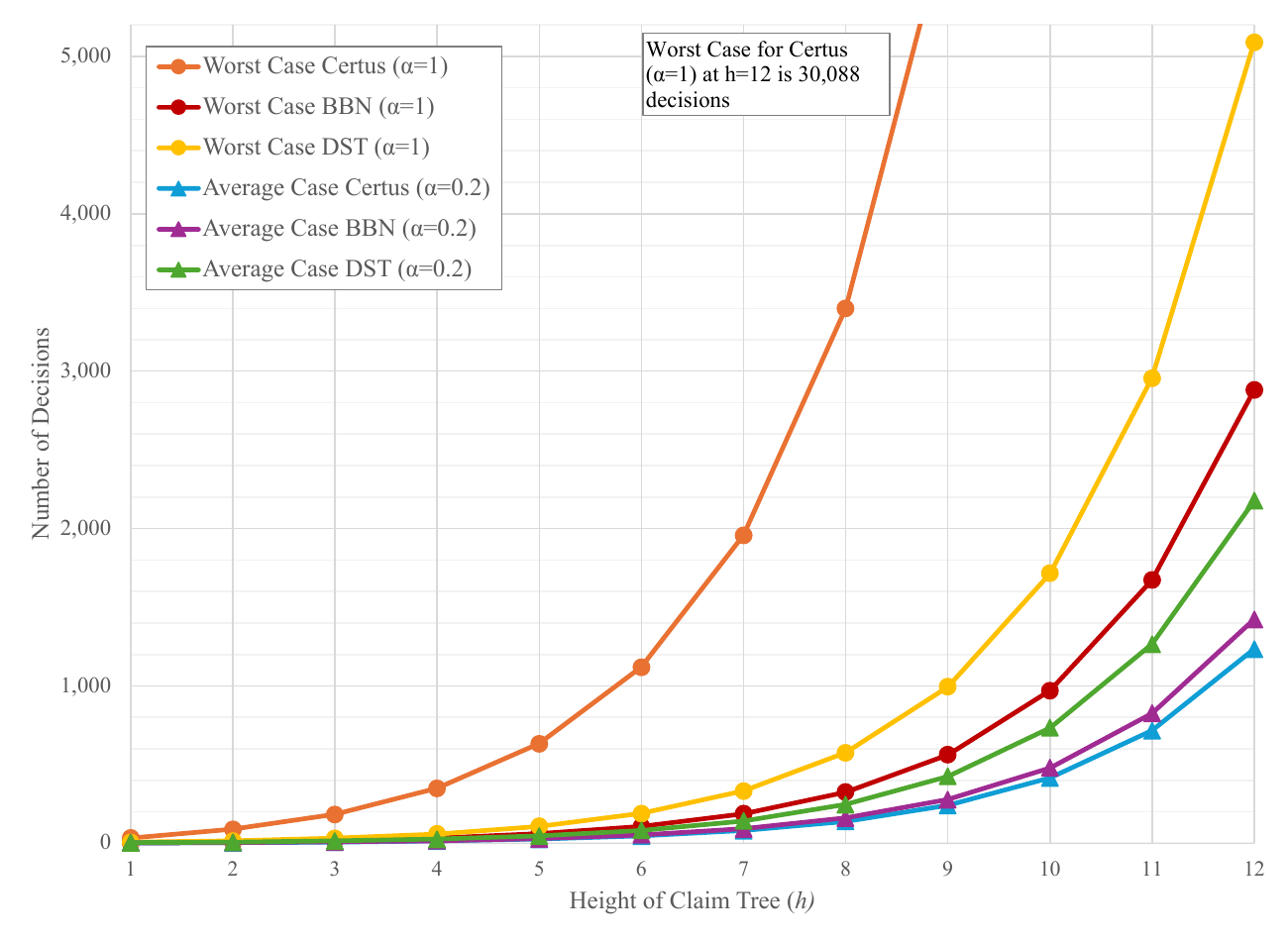}
    \caption{Scalability analysis results for $1 \leq h \leq 12$.}
    \label{fig:scalability-results}
\end{figure}

The worst and average case decision complexities for the \Certus{}, BBN, and DST methods are estimated using the model developed in Section \ref{sec:method} with the parameters selected in Section \ref{sec:param} for different choices of argument height (parameter $h$). The results are visualized in Figure~\ref{fig:scalability-results}. Table~\ref{tab:scalability-decisions} gives the detailed decision complexity values and Table~\ref{tab:scalability-time} gives the corresponding effort estimates. 

The results for $h=10$ are taken as representative in terms of argument size, based on published case studies in Table~\ref{tab:argument-sizes}. At this size, \Certus{} has the worst worst-case decision complexity, requiring over 10,000 decisions, compared to the BBN method (971 decisions) and the DST method (1,716 decisions). However, in the average case \Certus{}' decision complexity is the lowest. A similar trend exists for the effort estimates. On average, for an argument with approximately 350 nodes, our model predicts that it would take 13 hours of effort to apply \Certus{}, 14 hours for the BBN method, and 26 hours for the DST method. 

\begin{table}[htbp]
\centering
\footnotesize
\setlength{\tabcolsep}{3.5pt}
\caption{Decision complexity ($O_d$ / $\Omega_d$) in the worst and average cases.}
\label{tab:scalability-decisions}
\begin{tabular}{cccccccccc}
\toprule
\multicolumn{4}{c}{\multirow{2}{*}{\textbf{Argument Size}}} &
\multicolumn{3}{c}{\textbf{Worst Case ($O_d$, $\alpha = 1$)}} &
\multicolumn{3}{c}{\textbf{Average Case ($\Omega_d$, $\alpha = 0.20$)}} \\
\cmidrule(lr){5-7} \cmidrule(lr){8-10}
& & & &
\Certus{} & BBN & DST &
\Certus{} & BBN & DST \\
\midrule
\textbf{$h$} & \textbf{$|C|$} & \textbf{$|E|$} & \textbf{$|C|+|E|$} &
$O_d$ & $O_d$ & $O_d$ &
$\Omega_d$ & $\Omega_d$ & $\Omega_d$ \\
\midrule
1  & 1   & 1   & 2     & 34      & 4     & 7      & 2    & 2    & 4    \\
2  & 2   & 2   & 4     & 89      & 9     & 16     & 4    & 5    & 8    \\
3  & 4   & 3   & 7     & 185     & 18    & 32     & 8    & 9    & 14   \\
4  & 8   & 5   & 13    & 350     & 34    & 60     & 14   & 17   & 26   \\
5  & 14  & 9   & 23    & 633     & 61    & 108    & 25   & 30   & 47   \\
6  & 25  & 15  & 40    & 1,120   & 108   & 190    & 44   & 53   & 82   \\
7  & 44  & 26  & 69    & 1,958   & 188   & 332    & 76   & 93   & 143  \\
8  & 76  & 45  & 120   & 3,399   & 326   & 576    & 131  & 161  & 247  \\
9  & 131 & 77  & 207   & 5,878   & 563   & 995    & 226  & 278  & 426  \\
\rowcolor{rowhighlight}
10 & 226 & 132 & 357   & 10,141  & 971   & 1,716  & 390  & 480  & 735  \\
11 & 389 & 227 & 615   & 17,475  & 1,673 & 2,957  & 671  & 827  & 1,265 \\
\bottomrule
\end{tabular}
\end{table}

\begin{table}[htbp]
\centering
\footnotesize
\setlength{\tabcolsep}{4pt}
\caption{Effort estimates ($T$, in hours) in the worst and average cases.}
\label{tab:scalability-time}
\begin{tabular}{cccccccccc}
\toprule
\multicolumn{4}{c}{\multirow{2}{*}{\textbf{Argument Size}}} &
\multicolumn{3}{c}{\textbf{Worst Case ($T$, $\alpha = 1$)}} &
\multicolumn{3}{c}{\textbf{Average Case ($T$, $\alpha = 0.20$)}} \\
\cmidrule(lr){5-7} \cmidrule(lr){8-10}
& & & &
\Certus{} & BBN & DST &
\Certus{} & BBN & DST \\
\midrule
\textbf{$h$} & \textbf{$|C|$} & \textbf{$|E|$} & \textbf{$|C|+|E|$} &
$T$ & $T$ & $T$ &
$T$ & $T$ & $T$ \\
\midrule
1  & 1   & 1   & 2     & 1   & 0  & 0  & 0  & 0  & 0  \\
2  & 2   & 2   & 4     & 2   & 0  & 0  & 0  & 0  & 0  \\
3  & 4   & 3   & 7     & 5   & 0  & 1  & 0  & 0  & 1  \\
4  & 8   & 5   & 13    & 9   & 1  & 1  & 0  & 1  & 1  \\
5  & 14  & 9   & 23    & 17  & 1  & 2  & 1  & 1  & 2  \\
6  & 25  & 15  & 40    & 30  & 2  & 4  & 1  & 2  & 3  \\
7  & 44  & 26  & 69    & 52  & 4  & 7  & 3  & 3  & 5  \\
8  & 76  & 45  & 120   & 90  & 6  & 11 & 4  & 5  & 9  \\
9  & 131 & 77  & 207   & 156 & 10 & 20 & 8 & 8  & 15 \\
\rowcolor{rowhighlight}
10 & 226 & 132 & 357   & 270 & 18 & 34 & 13 & 14 & 26 \\
11 & 389 & 227 & 615   & 465 & 31 & 59 & 23 & 24 & 45 \\
\bottomrule
\end{tabular}
\end{table}

\section{Discussion}\label{sec:discussion}

From the scalability analysis results above, in the worst case, a user applying \Certus{} to assess belief in an \AC{} must make significantly more decisions than when using the BBN or DST methods. This is not surprising: among the methods analyzed, \Certus{} aims to offer greater flexibility in belief propagation expressions, which comes at the cost of higher decision complexity in extreme scenarios \cite{diemert2026thesis}. However, in the average case, \Certus{} users make marginally fewer decisions than BBN users and roughly half as many as DST users.

Beyond the results of the scalability analysis, this paper makes two conceptual contributions to the field of quantitative confidence assessment. First, in response to concerns raised by practitioners \cite{diemert2025cams}, it has introduced two measures to understand the additional work required to apply a \CAM{}: decision complexity (number of decisions) and level of effort (in hours). Second, it has proposed a numerical model for estimating these quantities. The application of the model to three different quantitative \CAM{}s serves as a means of preliminary validation of the concept.

There are several limitations and areas of future work that could be explored related to the effort required to apply quantitative \CAM{}s.

\tightParagraphSpacing{}
\paragraph{Limitations of the Argument Model.} The argument model is an idealized $n$-ary tree chosen for analytical tractability. Real \AC{}s are not uniform trees: some have long chains of reasoning that are considerably deeper than the average, and others have a handful of wide argument steps alongside mostly narrow ones. In the present study, this concern is partially mitigated by grounding parameter choices in published case studies, but the model is unlikely to perfectly predict decision complexities for real-world \AC{}s. A future analysis could use statistical simulations parameterised with structural data from industrial partners.

\tightParagraphSpacing{}
\paragraph{Limitations of the Decision Parameters.} The model counts the \textit{number} of decisions but not their relative \textit{difficulty}. For instance, the DST method requires $v=2$ decisions per leaf valuation on a five-point ordinal scale, whereas the BBN method requires one decision on a continuous $[0,1]$ scale. Whether one is harder than the other for \AC{} developers is an open empirical question not addressed by the current model.

\tightParagraphSpacing{}
\paragraph{Limitations of the Timing Parameters.} The timing parameters $t_v$ and $t_p$ were chosen based on intuition rather than empirical data, and are likely context-dependent (e.g., developer skill, argument domain). As a result, the effort estimates should be treated as rough illustrations rather than precise predictions. A calibration study would be necessary to produce accurate timing parameters for a given deployment context.

\begin{credits}
\subsubsection{\ackname} This work was supported by the Natural Sciences and Engineering Research Council (NSERC).

\subsubsection{\discintname}
In addition to his role at the University of Victoria as a Ph.D. student, author S. Diemert is employed by Critical Systems Labs Inc., a Vancouver-based company that develops commercial tools for \AC{} management.
\end{credits}

%
%
%
\bibliographystyle{splncs04}
\bibliography{refs}

\end{document}